\documentclass[a4paper,superscriptaddress,aip,apl,reprint]{revtex4-1}
\usepackage[T1]{fontenc}
\usepackage{graphicx}
\usepackage{amsmath}
\usepackage{dcolumn}
\usepackage{bm}
\usepackage{natbib}

\DeclareTextSymbol{\degre}{T1}{6}

\begin{document}

\title{Reduction of magnetostatic interactions in self-organized arrays of nickel nanowires using atomic layer deposition}

\author{S. Da Col}\email{sandrine.da-col@grenoble.cnrs.fr}
\affiliation{Institut N\'eel, CNRS et Universit\'e Joseph Fourier, BP 166, F-38042 Grenoble Cedex 9, France}
\author{M. Darques}
\affiliation{Institut N\'eel, CNRS et Universit\'e Joseph Fourier, BP 166, F-38042 Grenoble Cedex 9, France}
\author{O. Fruchart}
\affiliation{Institut N\'eel, CNRS et Universit\'e Joseph Fourier, BP 166, F-38042 Grenoble Cedex 9, France}
\author{L. Cagnon}
\affiliation{Institut N\'eel, CNRS et Universit\'e Joseph Fourier, BP 166, F-38042 Grenoble Cedex 9, France}

\date{\today}

\begin{abstract}
Ordered arrays of magnetic nanowires are commonly synthesized by electrodeposition in nanoporous alumina templates.
Due to their dense packing, strong magnetostatic interactions prevent the manipulation of wires individually.
Using atomic layer deposition we reduce the diameter of the pores prior to electrodeposition.
This reduces magnetostatic interactions, yielding fully remanent hysteresis loops.
This is a first step towards the use of such arrays for magnetic racetrack memories.
\end{abstract}

\maketitle


The progress of the areal density of magnetic recording has been exponential over half a century.
However it shall soon reach a halt because of thermal stability and considering as an ultimate bound that the grain size in current hard disk drives is not much larger than one order of magnitude the dimensions of atoms.
Securing progress on the long run requires a rupture in concepts, such as the three-dimensional magnetic racetrack memory, where bits are encoded in the form of magnetic domain walls along vertical magnetic nanowires arranged in dense arrays \cite{parkin_shiftable_2004}. 

Competitive third dimension racetrack memories would require dense arrays of nanowires with diameter a few tens of nanometers and length several tens of micrometers.
While these cannot be achieved by top-down techniques, self-organized nanoporous templates obtained by anodization\cite{masuda_ordered_1995} (possibly long-ranged ordered thanks to an initial step of lithography such as nanoimprint\cite{lee_fast_2006}) and filled by electrodeposition meet these requirements, and have been used and optimized for many years \cite{whitney_fabrication_1993, nielsch_hexagonally_2001}.
A bottleneck of this approach is that although anodization processes may be varied to tune the pitch $D$ of the array, the ratio of pore diameter $d$ with $D$ is essentially fixed and not much less than a fraction of unity \cite{lee_fast_2006}.
This yields dense arrays with strong magnetostatic interactions between nanowires, preventing their use as independent entities as required in memories.
At a macroscopic level, this manifests itself as a loss of remanence of the arrays, some wires reversing spontaneously under the demagnetizing influence of their neighbors.

It was confirmed experimentally that magnetostatic interactions are enhanced when the $d/D$ ratio increases, based on chemical etching of the pores after anodization \cite{vazquez_magnetic_2004}, which is consistent with models of demagnetizing coefficients \cite{encinas-oropesa_dipolar_2001}.
On the contrary, one would need to decrease the pore diameter after anodization, to decrease magnetostatic interactions to the required level.
For this purpose we have used atomic layer deposition (ALD), able to yield  conformal deposits even on large aspect ratio structures such as nanoporous materials \cite{sander_template-assisted_2004, ott_modification_1997, elam_conformal_2003}.
This allowed us to reduce pore diameters from 50\,nm down to below 20\,nm, keeping the initial array pitch $D=$105\,nm.
A dramatic decrease of magnetostatic interactions followed, which resulted in fully remanent hysteresis loops of the arrays.\\

\begin{figure}
  \begin{center}
  \centerline{\includegraphics [width=8.5cm]{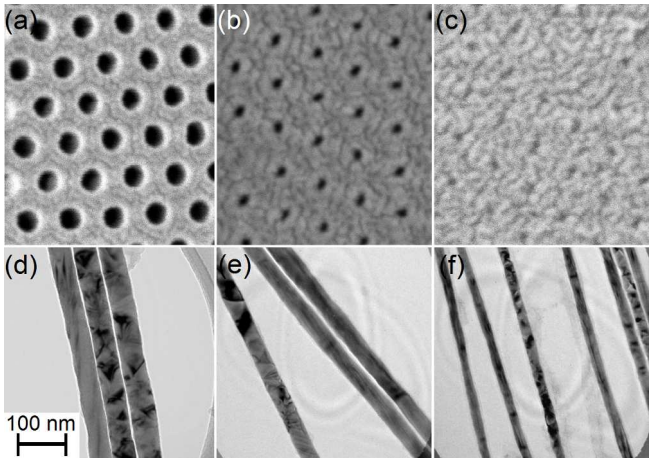}}
  \caption{Top view SEM image of empty alumina membranes (top) and TEM images of Ni nanowires (bottom) for (a,d) 0, (b,e) 100 and (c,f) 150 ALD cycles (same scale bar on all images).}
  \label{fig1}
  \end{center}
\end{figure}

Porous alumina membranes were prepared by a two-step anodization process of aluminum \cite{dahmane_magnetic_2006}.
We used 0.5\,M oxalic acid at 15\degre C, potential 40\,V during 17+7\,h, and a 30\,min pore opening step in phosphoric acid.
This yields pores organized over domains of lateral size several micrometers  with pore length $t=66\,\mu$m, pore diameter $d=46$\,nm and  interpore distance $D=105$\,nm (figure~\ref{fig1}.a).

The pore diameter was then reduced by depositing Al$_2$O$_3$ by ALD.
Contrary to chemical vapor deposition where all reactants necessary for growth are present simulaneously in the chamber, the ALD reaction is based on the saturating chemisorption of reactants sequentially injected in the deposition chamber along with a carrier gas (N$_2$).
This cyclic process allows for a precise control of the deposited thickness, as one cycle is responsible for the formation of a stoechiometric monolayer of material.
The saturated chemisorption allows one to perfectly match the substrate topography, however complex it may be.

The precursors used for alumina formation were trimethyl-aluminum and water, with a growth rate of 0.1\,nm per cycle \cite{puurunen_a-case_2005}.
The exposure time was set to 60\,s for each precursor for the homogeneous diffusion on the surfaces deep inside the pores.
Several parts of the same porous template were exposed to 0, 35, 75, 100 or 150 ALD cycles.
The empty membranes and wires released from the alumina were observed by scanning electron microscopy (SEM) and transmission electron microscopy (TEM), respectively (figure~\ref{fig1}).
The geometrical features of the array and nanowires after ALD are summarized in table \ref{tab1}.
All parameters are derived from SEM.
The decrease of pore diameter with the number of ALD cycles is consistent with the growth rate observed on flat surfaces.

 The electrodeposition was finally done for 1\,h at -1\,V in an acid electrolyte containing nickel ions after the backside was metallized by gold sputtering. 

\begin{table}
\caption{\label{tab1}Pores and nanowires geometry depending on the number of ALD cycles $n_{\mathrm{ALD}}$: length $L$, mean diameter $d$, aspect ratio and membrane's porosity $p$\cite{porosity}.}
  \begin{ruledtabular}
  \begin{tabular}{ccccc}
$n_{\mathrm{ALD}}$ & $L$ ($\mu$m) & $d$ (nm) & aspect ratio & $p$ ($\%$) \\
\hline
0 & 10.0 & 46 \(\pm\) 3 & 220 \(\pm\) 10  & 17.3 \(\pm\) 2.1  \\
35 & 13.0 & 39 \(\pm\) 4 & 330 \(\pm\) 30 & 12.7 \(\pm\) 2.6  \\
75 & 20.0 & 32 \(\pm\) 3 & 620 \(\pm\) 70 & 8.5 \(\pm\) 1.8  \\
100 & 19.0 & 27 \(\pm\) 3 & 700  \(\pm\) 90 & 6.0 \(\pm\) 1.3  \\
150 & 28.5 & 18 \(\pm\) 3 & 1900 \(\pm\) 290 & 2.7 \(\pm\) 1.0  \\
  \end{tabular}
  \end{ruledtabular}
\end{table}


Single wires are expected to be essentially single-domain, and reverse their magnetization through nucleation of a domain wall at one end, followed by its quick propagation along the wire length \cite{hertel_micromagnetic_2001}.
In an array these single-domain wires interact through dipolar fields.
The intrinsic switching field distribution (SFD) of an assembly of wires if considered non-interacting is widened: starting from saturation the internal dipolar field is globally demagnetizing inside the array and adds up to the applied field, resulting in the early switching of the first wires with respect to their intrinsic coercive field.
Similarly, when more than half of the wires have reversed the internal dipolar field in the array are stabilizing for the wires that have not yet reversed, delaying their reversal.

Magnetization reversal processes of the arrays were studied by magnetization loops measured by vibrating sample magnetometer (VSM) at room temperature.
We performed minor hysteresis loops, first to confirm the abrupt nucleation/propagation reversal mode, second to extract reliably the SFD.
For each recoil point the backward slope is associated with reversible processes, while its difference with the forwards slope yields the density of switching events.

\begin{figure}
  \begin{center}
  \centerline{\includegraphics [width=8.5cm]{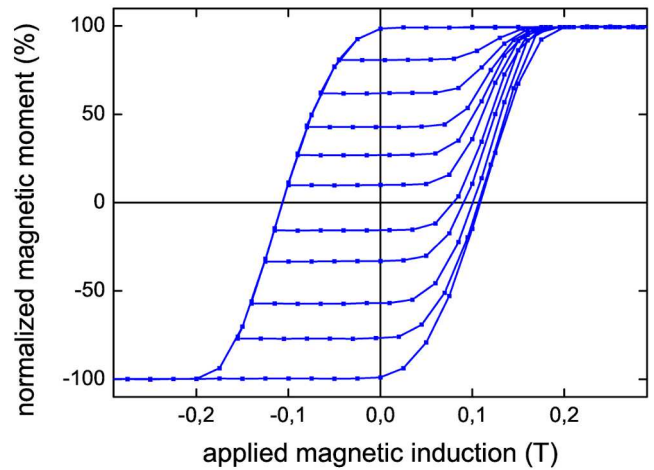}}
  \caption{Major and minor hysteresis loops measured by VSM at room temperature on the sample with $d=$27\,nm.}
  \label{fig2}
  \end{center}
\end{figure}
As shown in figure~\ref{fig2} (for $d=27$\,nm) the reversible contribution is nearly negligible.
Wires with all other diameters exhibit the same behaviour.
These results give a convincing basis to magnetization switching through nucleation and propagation.
The major loops thus result most entirely from switching events, so that the SFDs can safely be extracted as the normalized derivative of these major loops.
Only the major loop derivative is used hereafter.

\begin{figure}
  \begin{center}
  \centerline{\includegraphics [width=8.5cm]{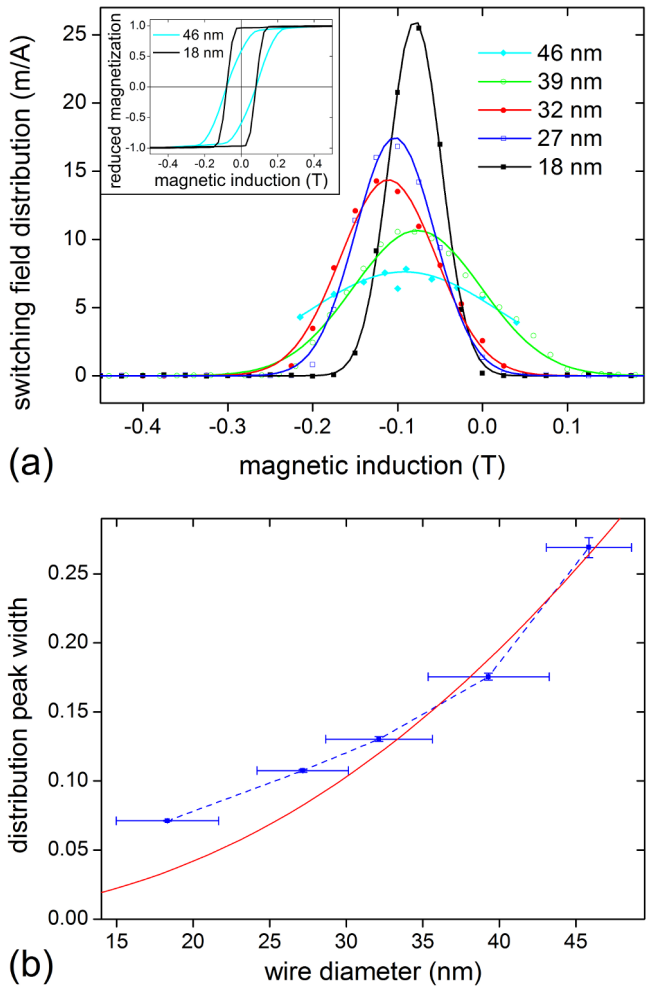}}
  \caption{(a) SFDs for different wire diameters determined from major hysteresis loops: Experimental values (symbols) are fitted with gaussian curves (solid lines) (b) Demagnetization field dependance with wire diameter: experiemental values (squares on dashed line) compared to the model (solid line).}
  \label{fig3}
  \end{center}
\end{figure}
The resulting SFDs coming from hysteresis loops measured for samples with wires of different diameters are shown in figure~\ref{fig3}(a).
The experimental values are fitted to a Gaussian curve centered around the coercive field.

The coercivity is only weakly depending on the wire diameter and follows a non-monotonic behavior, the reduction of diameter favoring higher reversal fields, counterbalanced at room temperature by increased thermal activation in low dimensions \cite{zeng_structure_2002}.
On the contrary the SFD is dramatically and monotonically reduced for smaller diameters, which we analyze quantitatively below.

Assuming wires homogeneously magnetized up to their very extremities, the stray field $H_d$ acting along a given wire may be calculated as arising from the so-called magnetic charges located at the ends of all other wires.
We use the mean-field approximation, where surfaces charges are related to the average moment of the array at any step of the reversal.
We consider a wire still not reversed, with magnetization up, and focus on the magnitude of $H_d$ in the vicinity of one end of the wire, where nucleation occurs (we consider the top end, without loss of generality).
The maximum strength of the internal dipolar field $H_d$ felt by a given wire is given by \cite{equation}:
\begin{eqnarray}
\label{tot}
\frac{H_d}{M_s}=&&-\frac{\pi}{4\sqrt{3}}\left(\frac{d}{ D}\right)^2
  -\eta\left(\frac{3}{8}+\frac{\pi}{6\sqrt 3}\right)\left(\frac{d}{D}\right)^3 \nonumber\\
&&+\eta^3\left(\frac{9}{16}+\frac{\pi}{27\sqrt3}\right) \left(\frac{d}{D}\right)^5.
\end{eqnarray}
where $M_s$ is the spontaneous magnetization and $\eta$ the normalized depth along the wire length.
For numerical application, we will use $\eta=1$. 
As our model is a refinement of that of ref.\cite{wang_in-field_2008} which was dealing with shorter wires, here we need to take into account the finite length of nucleation volumes, given the long length of the wires.

The experimental strength of dipolar fields is estimated as the central width of the Gaussian of figure~\ref{fig3}(a) encompassing 50\,\% of its area, which is $2\sqrt{2}\,\mathrm{erf}^{-1}(0.5)\,\sigma$, where $\sigma^2$ is the variance of the Gaussian.

The experimental and modeled strengths of dipolar fields are displayed in figure~\ref{fig3}(b).
They reveal a remarkable agreement, given the absence of adjustable parameter in the theory.
This also hints at a very low intrinsic SFD of the wires if not in interaction.\\


As a result the strong reduction of the SFD allows us to recover fully remanent hysteresis loops for the arrays (insert of figure~\ref{fig3}(a)), a principle that may be applied to any nanoporous template.
This lifts a long-standing bottleneck of anodized nanoporous templates, where magnetostatic interactions usually dictate collective magnetization processes.
This opens the way to the use of such nanowires still in their matrix for fundamental as well as applied purposes, where each nanowire may be addressed independently of the others.
In racetrack memories however, the issue is more severe because domain wall propagation fields would be much smaller than nucleation field (the coercivity in the present study), so that the interactions should be much more reduced. \\


J. Debray is thanked for his help in samples preparation.

\bibliographystyle{apsrev}

\begin{thebibliography}{16}
\bibitem{porosity} In hexagonal geometry the membrane porosity is $p=\frac{\pi d^2}{2\sqrt{3}D^2}$.
\bibitem{equation} Equation~\ref{tot} is a sum of 3 contributions.
\\First, the opposite side is far from the wire end and is thus seen as carrying an average charge density $-p\langle M \rangle$, giving rise to the first term in equation \ref{tot}.
\\Then, the nucleation side is treated in two parts, for the sake of accuracy, whose sum gives the second and third terms:
\\- the ring of first neighbors is treated with point surface charges located on the emerging axis of each neighboring wire, giving rise to a term
$-\frac{3}{8} \left(\frac{d}{D}\right)^3 \left[{1-\frac{3}{2}\eta^2\left(\frac{d}{D}\right)^2}\right]$;
\\- in the outer part starting at the center of second nearest neighbors, the charge density $p\langle M \rangle$ is modeled as homogeneous, giving rise to a term
$ -\frac{\pi}{6\sqrt{3}}\eta \left(\frac{d}{D}\right)^3 \left[{1-\frac{2}{9}\eta^2\left(\frac{d}{D}\right)^2}\right]$.\\
In both cases the resulting field is vertical ($z$ axis), and estimated at a reduced depth $\eta=z/d$, characterizing the center of the nucleation volume.
\end{thebibliography}

\end{document}